# Atomic to mesoscale hierarchical structures and magnetic states in an anisotropic layered ferromagnet FePd$_2$Te$_2$

Shuo Mi[1,2,+], Manyu Wang[1,2,+], Bingxian Shi[1,2,+], Songyang Li[1,2], Xiaoxiao Pei[3], Yanyan Geng[1,2], Shumin Meng[1,2], Rui Xu[1], Li Huang[3], Wei Ji[1,2], Fei Pang[1,2], Peng Cheng[1,2,*], Jianfeng Guo[3,*], Zhihai Cheng[1,2,*]

[1]*Key Laboratory of Quantum State Construction and Manipulation (Ministry of Education), School of Physics, Renmin University of China, Beijing 100872, China*

[2]*Beijing Key Laboratory of Optoelectronic Functional Materials & Micro-nano Devices, School of Physics, Renmin University of China, Beijing 100872, China*

[3]*Beijing National Laboratory for Condensed Matter Physics, Institute of Physics, Chinese Academy of Sciences, Beijing 100190, China*

**Abstract:** Two-dimensional (2D) magnetic materials have predominantly exhibited easy-axis or easy-plane anisotropy and display a high sensitivity to the underlying crystal structure and lattice symmetry. Recently, an in-plane anisotropic 2D ferromagnet of FePd$_2$Te$_2$ has been discovered with intriguing structure and quasi-one-dimensional spin system. Here, we report a real-space investigation of its twinning structure and magnetic states using atomic/magnetic force microscopy (AFM/MFM) combined with scanning tunneling microscopy (STM). The atomic to mesoscale hierarchical structures with the orthogonal and corrugated compressive /tensile(C/T) regions are directly observed due to the intrinsic twinning-domain characteristic. The structure-related intact ferromagnetic (FM), field-induced polarized-FM states and their transitions are comparatively discussed at the mesoscale with the corresponding macroscopic magnetic measurements. Temperature- and field-dependent evolution of magnetic phase are further investigated at the FM and PM states, and summarized to obtain a unique H-T phase diagram of FePd$_2$Te$_2$. Our work provides key results for understanding the complicated magnetic properties of FePd$_2$Te$_2$, and suggests new directions for manipulating magnetic states through the atomic and mesoscale structure engineering.

[+] These authors contributed equally to this work.

*Correspondence to Email: zhihaicheng@ruc.edu.cn; pcheng@ruc.edu.cn; jfguo@iphy.ac.cn

**Introduction**

Two-dimensional (2D) magnetic materials have attracted substantial interest due to their potential applications in spintronics, quantum computing, and information storage [1-6]. The realization of long-range magnetic order in atomically thin systems such as $CrI_3$ and $Fe_3GeTe_2$ has challenged conventional understanding of magnetism in low dimensions and triggered extensive investigations into the mechanisms stabilizing magnetic order at the atomic scale [7-9]. Among these mechanisms, Magnetic anisotropy plays a central role as it governs spin orientation and plays a vital role in counteracting thermal fluctuations, thereby enabling robust magnetism in the 2D limit [2,10-12]. In particular, 2D magnets with in-plane magnetic anisotropy have garnered increasing attention due to their potential for in-plane spin manipulation, domain wall motion, and spin transport phenomena [11,13-15]. Intrinsic magnetic anisotropy and spin dimensionality are critical in determining the emergent magnetic behaviors in two-dimensional systems.

Magnetism in 2D systems is also highly sensitive to the underlying crystal structure and lattice symmetry. Structural factors such as local strain, twinning-domains, and lattice defects can significantly modulate exchange interactions and magnetic anisotropy [16-27]. For instance, in $Fe_3GaTe_2$ [16], folding the nanosheets induces structural deformation that enables magnetic domain modulation, while in $Fe_4GeTe_2$ [22], subtle variations in lattice parameters can trigger spin reorientation, leading to unconventional domain walls and intricate magnetic domain textures. Despite these advancements, controlling magnetic domain through targeted structural engineering and real-space manipulation, especially when scaling from the atomic to the mesoscopic level, remains a significant challenge.

Among the limited set of known 2D magnets with in-plane anisotropy, $FePd_2Te_2$ stands out as a promising candidate due to its unusual structural and magnetic characteristics [27-29]. Unlike conventional 2D magnets, $FePd_2Te_2$ combines quasi-1D Fe zigzag chains within a 2D layered lattice, offering a hybrid dimensionality with spin systems and 2D magnetism. The presence of twinning-domains, where the Fe atomic arrangement rotates by 90° at the domain boundaries, adds an additional degree of complexity by locally modulating magnetic exchange interactions and anisotropy. This coupling between lattice distortions and spin configurations provides a rich platform for investigating how structural inhomogeneities affect magnetic ordering in 2D systems.

In this work, using AFM/MFM/STM combined with macroscopic magnetization measurements, we reveal intrinsic hierarchical twinning-domain structures in $FePd_2Te_2$, characterized by corrugated topography and strain-induced compressive (C) and tensile (T) regions. Structural domains induce orthogonal in-plane magnetic anisotropy at low temperatures, with field-driven evolution leading to a spin crossover and transition into a field-polarized FM state. Crucially, C/T regions remain distinguishable in this polarized state, demonstrating persistent strain-magnetism coupling. Complex domain evolution is further observed under varying temperature/field, including a unique polarized-PM state arising from structural corrugation and strain. These findings elucidate the critical interplay between hierarchical twinning, local strain, and magnetic anisotropy, highlighting strain engineering for tailoring low-dimensional magnetism.

# Results and discussion

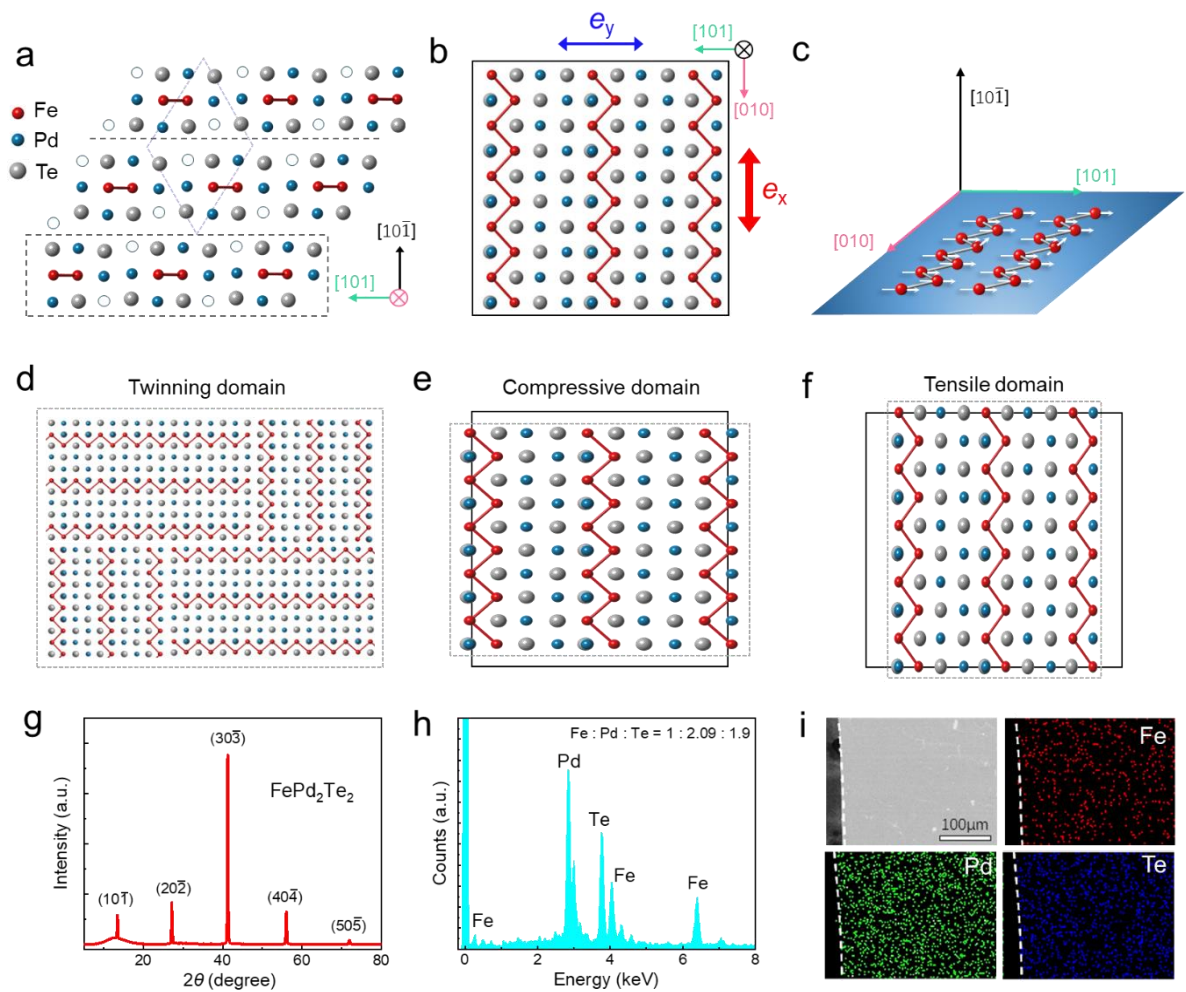

**Figure 1. Crystal structures of anisotropic layered FePd$_2$Te$_2$.** (a,b) Side view (a) and top view (b) of its atomic structure. The crystal structure shows a layered structure along the [10$\bar{1}$] directions. The crystal can be easily exfoliated between the Pd-Te layers, as indicated by the black dashed lines. The Fe-zigzag chains lie in the exfoliation (10$\bar{1}$) plane and along the [010] directions, and contribute anisotropic mechanical properties with large ($e_x$, along the chain) and small ($e_y$, vertical to the chain) Young's modulus. (c) Schematic diagram for the in-plane (vertical to the chain) spin orientations of Fe-chain atoms, indicated by the associated white arrows. (d) Schematic diagram of the coexistent two kinds of domains with orthogonal Fe-zigzag chains. (e,f) Schematics of the compressive (e) and tensile (f) domains after structural relaxation from the unstressed layers of (b) due to their distributed anisotropic domains. The "compressive" and "tensile" refer to lattice distortions along the direction of the Fe-zigzag chains. (g) XRD patterns from the cleavage plane of a FePd$_2$Te$_2$ single crystal. (h,i) EDS element analysis (h) and corresponding mapping (i) of one exfoliated FePd$_2$Te$_2$.

As a newly discovered layered magnetic material, single crystal of FePd$_2$Te$_2$ crystallizes in monoclinic symmetry with space group $P2_1/m$. The unit cell is illustrated by the lines in the side view of FePd$_2$Te$_2$ (Figure 1a) with the refined lattice parameters $a$ = 7.5024(5) Å, $b$ = 3.9534(2) Å, $c$ = 7.7366(7) Å, $α = γ = 90°$, and $β = 118.15°$ [29]. The cleavage (10$\bar{1}$) plane is highlighted by the dotted lines between the top and bottom Pd-Te sublayers of FePd$_2$Te$_2$ triple layers. The triple layer of FePd$_2$Te$_2$ features the anisotropic Fe-zigzag chains sandwiched between the top and bottom Pd-Te sublayers, as shown in Figure 1b, which is the origin of their complicated hierarchical structures and magnetic properties. Due to the crystal structure of FePd$_2$Te$_2$, atomic bonding is expected to be stronger along the chain direction ($e_x$, parallel to the [010] direction), leading to a relatively large Young's modulus [30,31]. In contrast, the bonding perpendicular to the chains ($e_y$, along the [101] direction), which involves weaker interlayer van der Waals interactions, results in a much smaller Young's modulus, as illustrated in Figure 1b. The Fe-zigzag chains in the middle sublayer thus govern both the anisotropic mechanical properties and magnetic characteristics, with an in-plane magnetic easy axis oriented along the [101] direction (Figure 1c).

As schematically shown in Figure 1d, the anisotropic triple-sublayer structure of FePd$_2$Te$_2$ has an intrinsic twinning-domain characteristic due to the quasi-1D arrangement of Fe-zigzag chains within the square-latticed Pd-Te matrix. Instead of forming a uniform single-domain structure, FePd$_2$Te$_2$ naturally develops orthogonal twinning-domains with Fe chains rotated by 90°, preserving the fourfold symmetry of the Te sublattice. This orthogonal twinning-domains induces anisotropic internal stress, leading to coexisting C and T regions in FePd$_2$Te$_2$, as illustrated in Figure 1e,f. These stress fields arise from the strong intrachain bonding and weak interlayer coupling of the quasi-1D Fe chains. Due to the crystallographic anisotropy in Young's modulus (Figure 1b), the C and T regions exhibit structurally distinct responses, introducing additional asymmetry into FePd$_2$Te$_2$. These structural inhomogeneities are not only key to the material's hierarchical domain architecture but may also significantly influence its magnetic properties from atomic to mesoscale, potentially involving fragmented Fe chains [32]. At the atomic scale, local strain can modulate the effective magnetic moments of Fe atoms and alter magnetic exchange interactions within each domain [24-27]. Furthermore, the strong in-plane magnetic anisotropy associated with the Fe-zigzag chains can stabilize distinct spin configurations within individual twinning-domains and lead to orthogonal magnetization orientations at the mesoscale.

Single crystals of FePd$_2$Te$_2$ were grown by melting stoichiometric elements and characterized by XRD and EDS measurements. The XRD data in Figure 1g confirms the exfoliated (10$\bar{1}$) plane of the synthesized crystal. The further EDS elemental analysis (Figure 1h) and mapping (Figure 1i) confirm its chemical composition of FePd$_2$Te$_2$ with a uniform spatial distribution. The successful growth of high-quality single crystals lays a solid foundation for subsequent real-space investigations of the material's structural domains and magnetic properties.

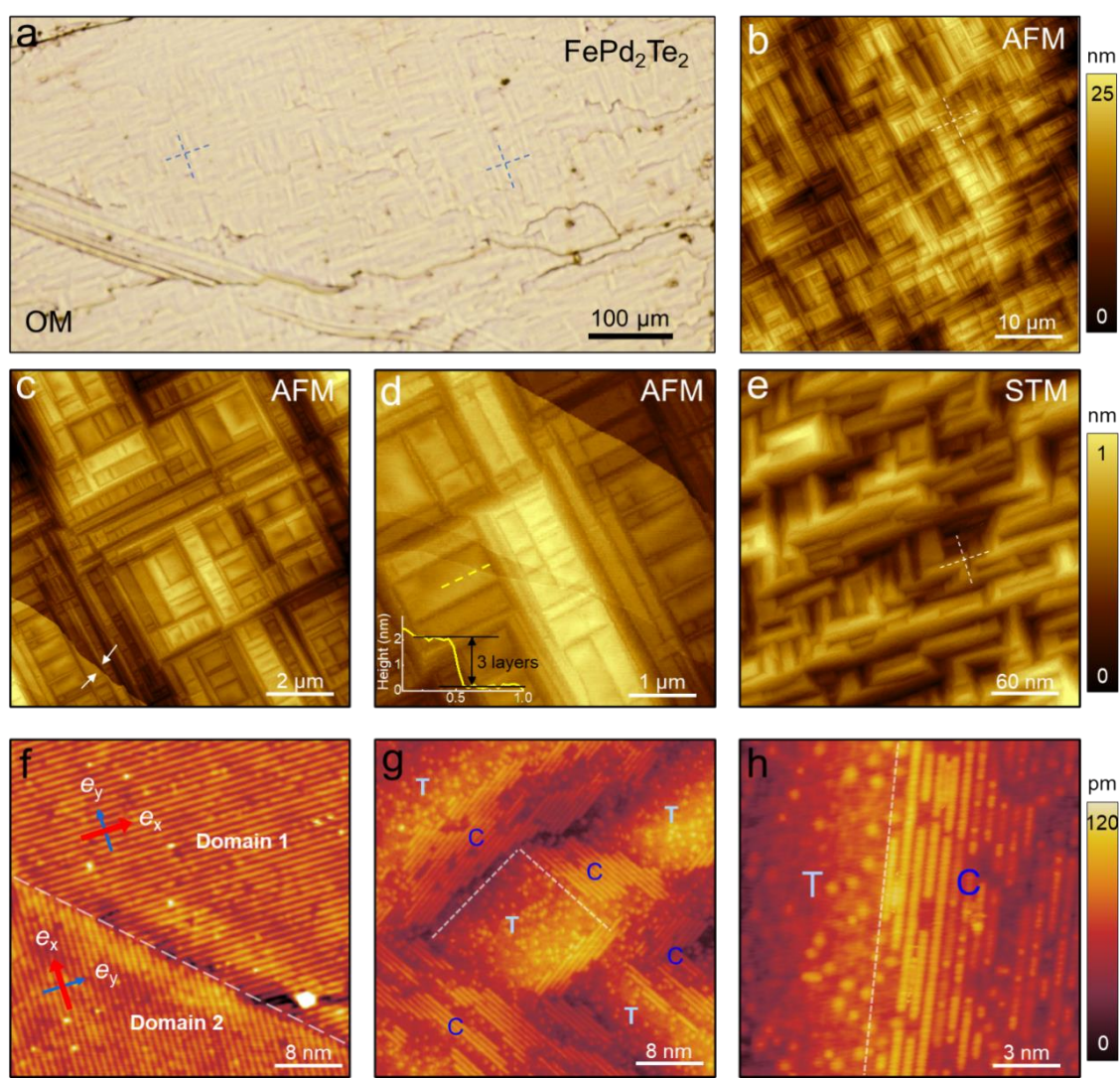

**Figure 2. Mesoscale AFM topography and atomic STM structure measurements of FePd$_2$Te$_2$.** (a,b) Optical microscopy (a) and large-scale AFM topography images of the fresh-cleaved crystal sample with hierarchical orthogonal characteristics. (c,d) Typical AFM topography images of the hierarchical corrugated regions from their intrinsic anisotropic twinning-domains. (e) Large-scale STM image with hierarchical orthogonal characteristics. (f) STM image of the pristine surface with two stitched twinning-domains of orthogonal Fe chains. The domain boundary is highlighted by the dashed white line. The anisotropic mechanical properties are represented by its different Young's modulus (large $e_x$, along the chain; small $e_y$, vertical to the chain). (g) STM image of the C and T domains, indicated by their relative stable (intact Fe chains) and unstable (broken Fe chains) properties after a short-time thermal treatment. (h) High-resolution STM image for the intact and broken Fe chains of the surface layer at the C and T domains.

Figure 2a exhibits the typical optical microscope image of $FePd_2Te_2$ crystal with complex orthogonal stripes, which originate from its intrinsic twinning-domain effect. The sizes of these orthogonal stripes are at the tens of micrometer, though more detailed features cannot be readily resolved at this scale. Figure 2b displays a large-scale AFM topography image of these stripes, in which the intricate hierarchical structures emerge clearly. These hierarchical structures are step-by-step made of orthogonal and corrugated regions to form this fractal-like structures at the small (hundreds of nanometers) to large (tens of micrometers) scale. Notably, these hierarchical features are consistently observed across different samples and persist after repeated cleaving, indicating that they are intrinsic rather than artifacts from surface cleavage.

Figure 2c,d shows the medium-scale AFM images of these hierarchical structures, in which the step edges (highlighted by the white arrows) are observed consistent with their layered structural characteristic. Interestingly, the layers exhibit a periodic corrugated topography that extends through adjacent layers, a rare characteristic among layered materials. These corrugations with vertical heights of ~10 nm (see Supplementary Figure S1), are attributed to structural relaxations arising from the anisotropic stress fields within distributed orthogonal twinning-domains, governed by the Fe-zigzag chains.

To further elucidate the twinning-domain structure at the atomic scale, STM was conducted at ~10 K. A representative large-area STM image is displayed in Figure 2e, where the hierarchical surface features are fully resolved. Figure 2f presents the high-resolution STM topography of a freshly cleaved surface at liquid nitrogen temperature, clearly revealing two orthogonal twinning domains distinguished by their Fe-zigzag chain orientations. A well-defined twin boundary separates these domains, emphasizing their structural distinction. Additional unprocessed STM images showing similar corrugated twinning-domain features are provided in Figure S2.

Upon a short-time thermal treatment, the twinning-domains can be further differentiated into C and T domains, as shown in the same scale STM image of Figure 2g. These C and T domains are identified based on the preservation or disruption of Fe chains at the surface layer, reflecting their differing structural stabilities [32-35]. The domain boundaries between C and T regions shows topographic features such as convex ridges or concave valleys. A representative convex ridge domain boundary is highlighted in Figure 2h, separating a C region with intact Fe chains from a T region with disrupted chains (see also Supplementary Figure S3).

These observations collectively demonstrate that $FePd_2Te_2$ shows a rich hierarchy of structural features extending from the atomic to the mesoscale. The corrugated surface topography, the emergence of C and T twinning-domains, and the formation of distinct domain boundaries such as convex ridges and concave valleys all stem from its intrinsic anisotropy and twinning-domain architecture. These structural inhomogeneities are not merely topographical, they are expected to play a significant role in modulating the spin configurations and governing the material's magnetic behavior [24-27,35,36]. To further explore these effects, high-sensitivity MFM measurements should be performed in real space, complemented by conventional magnetic susceptibility characterizations.

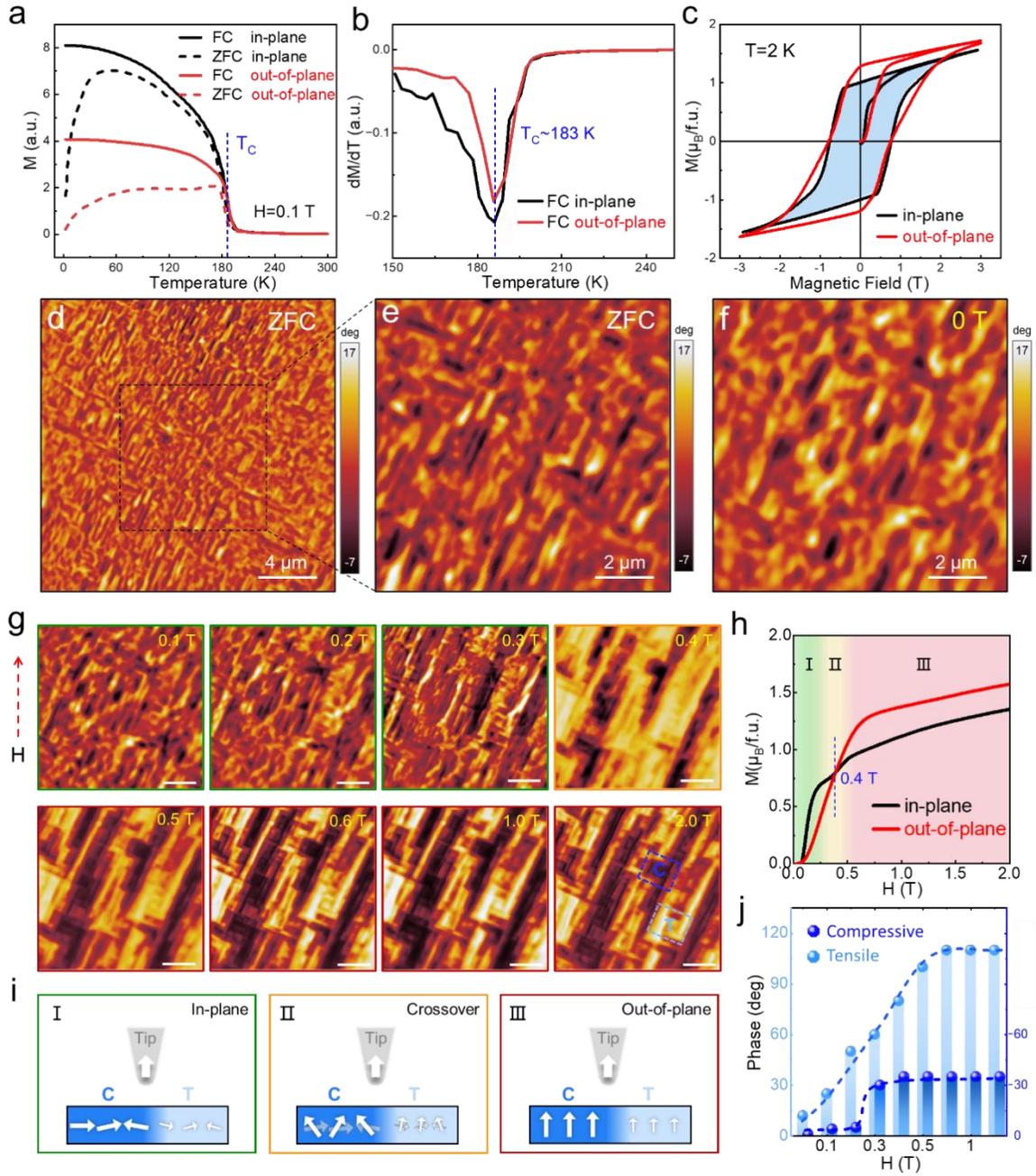

**Figure 3. Magnetic properties and MFM measurements of FePd$_2$Te$_2$.** (a) Temperature-dependent magnetic moments (M-T) with the out-of-plane (red) and in-plane (black) magnetic fields for the FC (solid line) and ZFC (dashed line) process at 0.1 T. (b) The derivative d*M*/d*T* curves from (a). (c) Magnetic hysteresis of out of-plane (red) and in-plane (black) at 2 K. (d,e) Typical MFM images at ~2 K after zero-field-cooling (ZFC) at large (d) and small (e) scale. (f) Typical MFM images at ~2 K after ZFC at 0 T. (g) Field-dependent MFM images of (f) with increased out-of-plane fields at three specific ranges of 0.1-0.3 T (I), ~0.4 T (II) and 0.4-2.0 T (III). (h) Initial M-H curves of out of-plane (red) and in-plane (black) at 2 K with three highlighted ranges (I, II and III). A clear crossover of in-plane to out-of-plane field-polarization is observed at ~ 0.4 T. (i) Schematic diagrams of the compressive (C, dark blue) and tensile (T, light blue) domain with large (big arrows) and reduced (small arrows) magnetic moments, respectively, at the three

field-polarization ranges (I, II and III). (j) Field-dependent MFM phase shifts of compressive (C, dark blue) and tensile (T, light blue) regions deduced from (g). The phase extraction procedure is detailed in Supplementary Figure S4.

Figure 3a shows the temperature-dependent magnetization curves measured under a 0.1 T magnetic field for both in-plane (black) and out-of-plane (red) directions. A clear paramagnetic-ferromagnetic (PM-FM) transition occurs around 180 K, with noticeably stronger magnetization observed along the in-plane direction, indicating a ferromagnetic order with enhanced in-plane magnetic anisotropy. The transition temperature is more precisely determined to be $T_C$ = 183 K without observable orientation-dependence, as shown by the dM/dT data in Figure 3b. To further evaluate magnetic anisotropy, Figure 3c shows the magnetic hysteresis (M-H) loops measured at 2 K for both in-plane (black lines) and out-of-plane (red lines). The in-plane direction exhibits a lower saturation field, indicating that $FePd_2Te_2$ is more easily magnetized along this direction [37]. The large hysteresis loops in both directions suggest that magnetic domains are stabilized in both in-plane and out-of-plane orientations, indicative of hard magnetic behavior [38].

To further investigate the real-space magnetic domain structure of $FePd_2Te_2$, MFM measurements were conducted after zero-field cooling (ZFC). A representative large-area MFM image in Figure 3d shows well-defined orthogonal contrast features, which are attributed to hierarchical structure-coupled magnetic domains. Higher-resolution MFM image (Figure 3e) reveals magnetic domains much finer than the underlying structural domains, suggesting a more complex spin configuration. The observed domain implies the coexistence of both in-plane and out-of-plane spin components.

The evolution of magnetic domains under increasing out-of-plane magnetic fields was further investigated by sequential MFM images after ZFC (see also Supplementary Figure S5). Figure 3f,g shows the gradual domain evolution as the field increases from 0 to 2 T, starting from the initial ZFC domain shown in Figure 3f and can be divided into three specific ranges. At 0-0.3 T, the in-plane spin moments within twinning-domains begin to flop toward the out-of-plane direction, and structural features become progressively visible in the MFM images [39,40]. At ~0.4 T, A spin-flop crossover of in-plane to out-of-plane occurs, marked by the sudden emergence of new domain contrast, aligned with the underlying orthogonal structural domains observed in the topography. At 0.5-2 T, the MFM contrast gradually saturates and remains stable, indicating the stabilization of an out-of-plane magnetized configuration. These three specific ranges correspond well with the field-dependent features observed in the initial M-H curves shown in Figure 3h.

Magnetic contrast persists under the saturation field, indicative of a polarized-FM state. This behavior is attributed to the unequal magnetic moments in the C and T regions (illustrated by the large and small arrows in Figure 3i), where the reduced moment in the T regions may result from Fe-chain fragmentation caused by tensile strain [24-27]. Based on the entire magnetic field evolution process, we present a schematic diagram illustrating the evolution of spin configurations in different strain regions, as shown in Figure 3i. Figure 3j further illustrates the evolution of magnetic domain within the C and T regions under increased magnetic fields. Interestingly, the Fe moments exhibit a gradual spin-flop transition in the T regions, whereas the C regions display an abrupt reorientation characteristic like the spin-flip transition [4,5], reflecting distinct magnetic responses between twinning-domain. Notably, the C regions reach magnetic saturation at a lower field than the T regions. These results demonstrate that the twinning-domain structure in anisotropic $FePd_2Te_2$ not only defines its hierarchical topography but also plays a key role in governing its magnetic domain behavior, enabling complex field-dependent spin reconfiguration through the interplay between C and T lattice distortions.

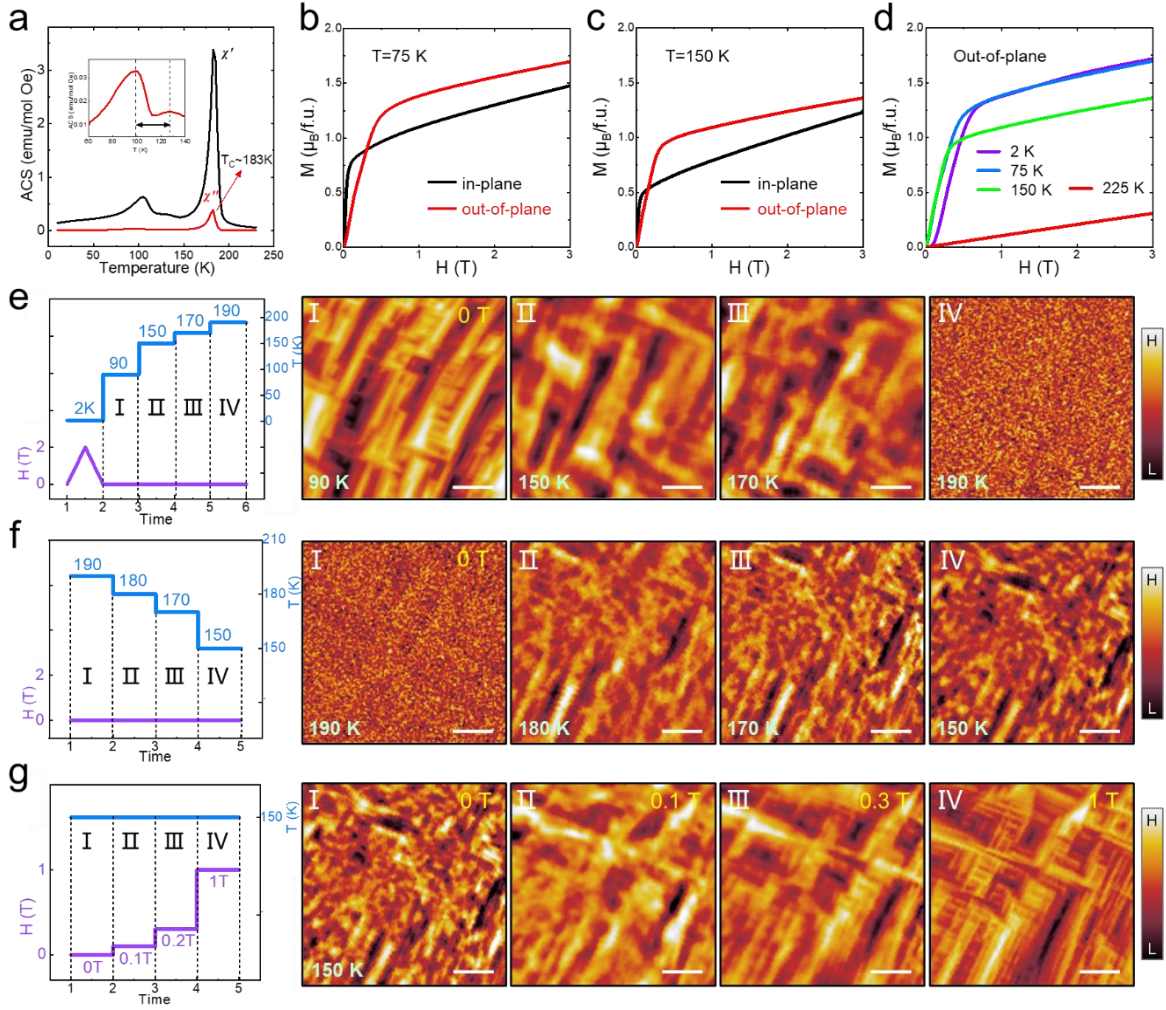

**Figure 4. Temperature-dependent magnetic susceptibility and MFM measurements of FePd$_2$Te$_2$.** (a) Temperature-dependent AC magnetic susceptibility measured under the magnetic field, indicating a clear PM-FM transition and broad magnetic crossover at ~183 K. The inset shows an enlarged view of $\chi_{ac}''(T)$ at ~100-120 K. (b,c) M-H curves of the out-of-plane (red) and in-plane (black) magnetizations at 75 K (b) and 150 K (c). (d) M-H curves of the out-of-plane magnetization at different temperatures. (e) Temperature-dependent MFM images from polarized-FM to PM state during zero-field warming process. (f) Temperature-dependent MFM images from PM to FM state during zero-field cooling process. (g) Field-dependent MFM images from FM to polarized-FM state at ~150 K.

To elucidate the temperature-dependent magnetic behavior of FePd$_2$Te$_2$, we performed systematic AC magnetic susceptibility (ACS) measurements and in-situ magnetic force microscopy (MFM) imaging. As shown in Figure 4a, ACS measurements exhibit a sharp peak at ~183 K, marking the PM-FM transition. Additionally, a broad feature in the 100-120 K range, which is absent in the static susceptibility (Figure 3a) and may indicate domain evolution driven by competing thermal fluctuations and magnetic anisotropy [41]. Complementary initial

M-H curves measured under out-of-plane and in-plane fields at 75 K and 150 K (Figure 4b,c) reveal distinct temperature-dependent magnetic anisotropy. With increasing temperature, the distinction between in-plane and out-of-plane responses progressively diminishes, beginning to approach isotropic magnetic behavior. The temperature evolution of the out-of-plane M-H curves from 2 K to 225 K reveals an enhancement of out-of-plane magnetic anisotropy in $FePd_2Te_2$ (Figure 4d).

To directly visualize the magnetic domain evolution in $FePd_2Te_2$, MFM measurements were performed under three distinct processes (Figures 4e-g). Starting from the polarized-FM state at 2 K, stepwise warming reveals that magnetic contrast weakens with increasing temperature and vanishes near 190 K (Figure 4e; Supplementary Figure S6), consistent with the sharp ACS peak at ~183 K marking the loss of long-range ferromagnetic order. Notably, distinct magnetic features persist up to ~120 K correlates well with the broad ACS feature in the 100-120 K range and above which the contrast becomes increasingly blurred, which may arise from the competition between thermal fluctuations and magnetic anisotropy. Subsequently, ZFC process across the PM-FM transition shows the magnetic contrast reappeared below ~183 K (Figure 4f; Supplementary Figure S7), gradually revealing domain features with orthogonal in-plane magnetic domain characteristics consistent with those at 2 K (Figures 3d-f), confirming that the PM-FM transition leads to the formation of structure-coupled in-plane magnetic domains. Comparison between ZFC and FC (Supplementary Figure S8) further demonstrates that applied fields promote spin alignment, facilitating the formation of the polarized-FM state. To investigate the temperature dependence of spin alignment dynamics and specifically to compare domain evolution before and after the fluctuation regime (~100-120 K), we examined field-induced domain changes at 150 K. As shown in Figure 4g (see also Supplementary Figure S9), increasing out-of-plane magnetic fields at 150 K induce a more rapid transition into the polarized-FM state than at low temperature, indicating reduced anisotropy energy at higher temperatures and easier spin alignment along the field direction.

Overall, the ACS and MFM images provide a coherent process of the magnetic evolution in $FePd_2Te_2$. From ~120 to ~180 K, long-range ferromagnetic order exists with relatively weak anisotropy, enabling easy spin polarization by external fields. Near 100-120 K, thermal energy rivals the anisotropy barrier, enhancing spin fluctuations and causing partially reoriented or dynamic domains [42-45]. Below ~100 K, increased magnetic anisotropy, likely due to coupling with structural domains, requires stronger fields to reorient spins. These findings highlight the temperature-dependent tuning of magnetic anisotropy and spin textures in FePd$_2$Te$_2$.

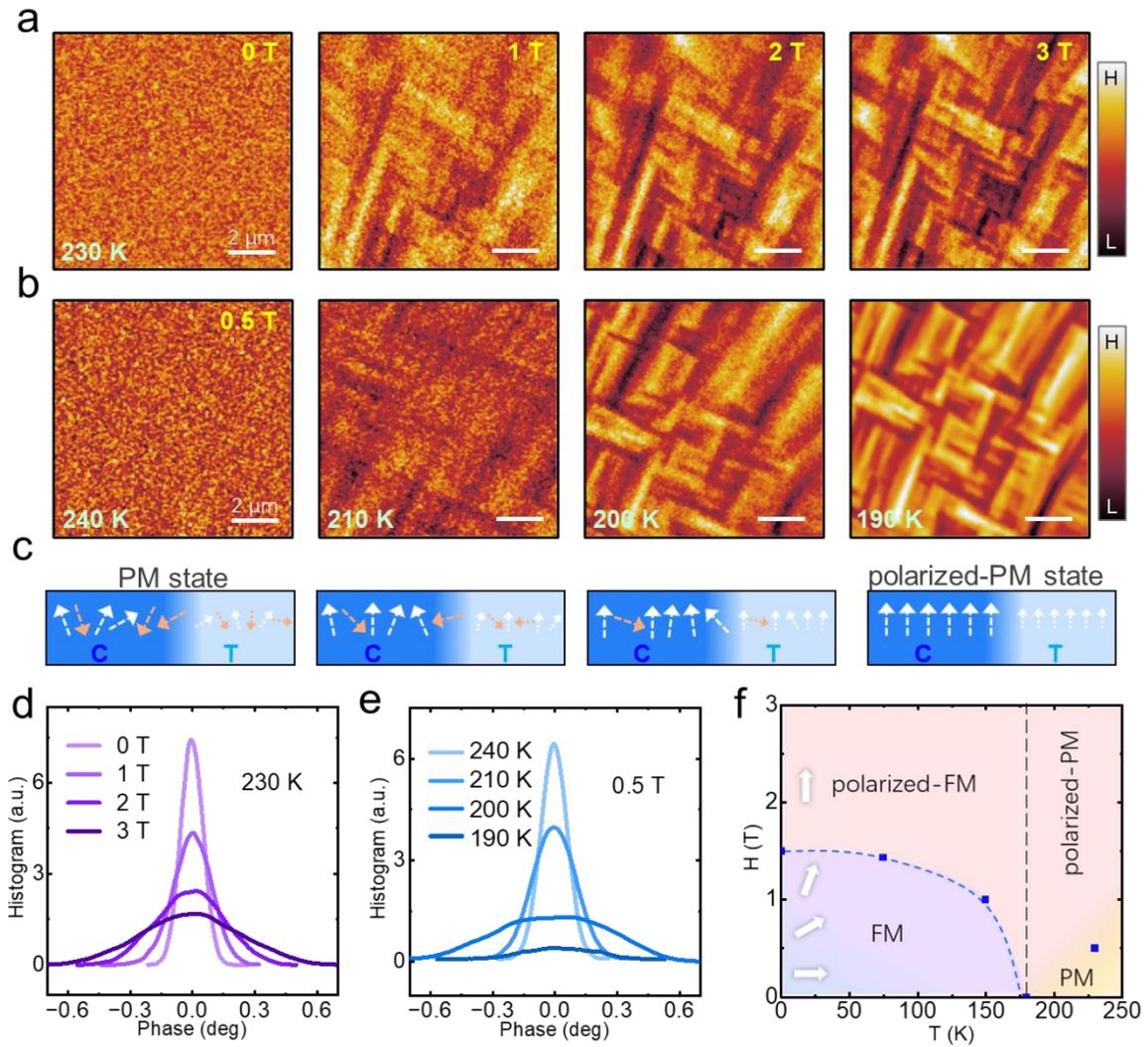

**Figure 5. Field-polarized spin textures at the PM state of FePd$_2$Te$_2$.** (a) Field-dependent MFM images from PM to polarized-PM state at 230 K. (b) Temperature-dependent MFM images after field cooling under 0.5 T, illustrating the transition from PM to polarized-FM state. (c) Schematic diagrams of polarized-PM state with the field-polarized magnetic moments of Fe atoms at the compressive (C, dark blue) and tensile (T, light blue) regions, respectively. (d) Histogram of MFM images in (a). (e) Histogram of MFM images of (b) at different temperatures under the magnetic field of 0.5 T. (d,e) Histograms of MFM images in (a) and (b), respectively. (f) H-T phase diagram of FePd$_2$Te$_2$, showing the FM (purple), polarized-FM (pink), and PM/polarized-PM state (gradient pink).

In the PM state, magnetic materials typically lack spontaneous long-range magnetic order [46,47]. However, in FePd$_2$Te$_2$, structure-related magnetic contrast is observed even in the PM state (Figure 5a,b; Supplementary Figures S10 and S11), arising from the intrinsic difference in magnetic moments between the C and T regions. At 230 K, the magnetic contrasts are gradually visualized with the increased magnetic fields, which is similar with the polarized-FM states (below $T_c$) and can be defined as a polarized-PM state. A similar evolution occurs under 0.5 T FC from 240 K to 190 K (Figure 5a), where structure-related magnetic domains gradually develop same as MFM images in Figure 5a. Figure 5c schematically illustrates the magnetic moment's behaviors observed in Figure 5a,b, where similar domain features emerge either by increasing magnetic field or decreasing temperature in PM state. This behavior can be semi-quantitatively described by the Boltzmann distribution, reflecting the competition between thermal fluctuations, which randomize spin orientations, and Zeeman energy, which promotes spin alignment along the magnetic field.

Figures 5d,e further supports this interpretation by showing histograms of the MFM phase extracted from the corresponding images in Figures 5a,b, respectively, highlighting the same behavior for gradual emergence of magnetic contrast under increased fields or decreased temperatures. The comprehensive magnetic phase diagram in Figure 5f summarizes the various magnetic states observed in FePd$_2$Te$_2$, including the complex FM state, the polarized-FM state, and the polarized-PM state. All these unique magnetic states are originated from the hierarchical twinning-domain structures of anisotropic FePd$_2$Te$_2$ from atomic to mesoscale levels, and indicate the potential possibility to further modulate its magnetic properties by delicately regulating their crystal structures. However, a quantitative theoretical understanding of how hierarchical structural features influence magnetic domain configurations remains a challenge. Further theoretical investigations are needed to elucidate the relationship between structural heterogeneity and magnetic domain behavior in these systems.

In summary, we have revealed the atomic-to-mesoscale hierarchical structure and its strong correlation with magnetic properties in the layered anisotropic magnet FePd$_2$Te$_2$. Using AFM/MFM and STM, twinning-domain induced C and T regions that modulate local spin configurations are observed, giving rise to distinct magnetic phases including FM and polarized-FM states. Temperature- and field-dependent measurements uncover a magnetic phase evolution marked by a crossover in anisotropy and domain dynamics, which are summarized in a detailed H-T phase diagram. Our work not only provide a unique material platform for the tunable magnetic properties, and will inspire more research interest in construct exotic magnetic states with the engineered material structures from atomic to mesoscale.

**Acknowledgments**

This project is supported by the National Key Research and Development Program of China (No. 2023YFA1406500), National Natural Science Foundation of China (NSFC) (No. 92477128, 12374200, 12074426, 12474148), and Fundamental Research Funds for the Central Universities and Research Funds of Renmin University of China (No. 21XNLG27). This paper is an outcome of "Study of Exotic Fractional Magnetization Plateau Phase Transitions and States in Low-dimensional Frustrated Quantum Systems" (RUC24QSDL039), funded by the "Qiushi Academic - Dongliang" Talent Cultivation Project at Renmin University of China in 2024.

## Materials and Methods

**Sample preparation and characterization.**

Single crystals of $FePd_2Te_2$ were grown by melting stoichiometric elements. Iron, palladium, and tellurium powder were mixed and ground in a molar ratio of 1:2:2. Deviation from this ratio would lead to the unsuccessful growth of large single crystals or the target phase. Then the mixtures were placed in an alumina crucible and sealed in a quartz tube under vacuum conditions. The entire tube was heated in a box furnace to 800 °C and held at that temperature for 2 days. Then it was cooled to 600 °C at a rate of 2 °C/h followed by annealing at this temperature for 2 days before being furnace-cooled to room temperature. In addition, we found that quenching the samples at 600 °C would improve the crystal quality, as revealed from X-ray characterization. However, direct quenching in the initial growth process seems to break the large crystal into small pieces. By reannealing and quenching the large single crystal grown by the initial furnace-cooled method, one could obtain crystals with both large size and good quality.

**Structure and Composition Characterization.**

The structure of $FePd_2Te_2$ crystals was analyzed at 273 K by XRD (D8 ADVANCE, Bruker) equipped with multilayer mirror monochromatized Mo K$\alpha$ ($\lambda$ = 0.71073 Å) radiation. The elemental composition and distribution were evaluated by EDS (X-MaxN 50 mm$^2$, Oxford Instruments) in the SEM.

**Magnetic Property Measurements.**

The measurements of magnetic susceptibility and $T_C$ were performed in a magnetic property measurement system (MPMS3, Quantum Design), in which the anisotropic magnetic properties of samples were observed separately. The temperature-dependent magnetic susceptibility for out-of-plane and in-plane magnetic fields was measured within the temperature range from 1.8 to 300 K by the processes of ZFC and FC with a field of 0.1 T, respectively. The field-dependent magnetization studies were performed with applied field range from 0 to 7 T at temperatures of 2, 75, 150, and 225 K for out-of-plane and in-plane in several. Besides, the measurement was performed in DC mode, in which the samples were scanned vertically 30 mm in 4 s, and the gradients of magnetic field and temperature were set as 100 Oe s$^{-1}$ and 2 K min$^{-1}$ with the same intervals of 100 Oe and 0.2 K, respectively.

**AFM and MFM measurements**

The AFM experiments were performed using a commercial atomic force microscope (Park, NX10) equipped with a commercial topography tip (Nanosensors, AC160TS, Quality factor ~500 at room temperature). The scanning probe system was operated at the resonance frequency of the topography tip, ~301 kHz. The AFM images were acquired in non-contact mode.

The MFM experiments were conducted using a commercial magnetic force microscope (attoAFM I, attocube) employing a commercial magnetic tip (Nanosensors, PPP-MFMR, Quality factor ~ 1800 at 2 K) based on a closed-cycle He cryostat (attoDRY2100, attocube). The scanning probe system was operated at the resonance frequency, ~ 75 kHz, of the magnetic tip. The MFM images were captured in a constant height mode with the scanning plane nominally ~100 nm above the sample surface. The MFM signal, i.e., the change in the cantilever phase, was proportional to the out-of-plane stray field gradient. The dark (bright) regions in the MFM images represented strong (weak) attractive magnetization, corresponding to regions with larger (smaller) magnetic moments. The MFM signal primarily reflects magnetic structures near the sample surface rather than the topmost atomic layer, as it originates from the stray field extending above the surface. Owing to the gradual spatial decay of magnetic dipolar interactions (~$1/r^2$), the tip can sense contributions from a subsurface volume. With a lift height of ~100 nm, the signal mainly represents magnetic domains within approximately the top 100 nm of the sample [4].

**STM characterization**

The samples were cleaved at room temperature and LN$_2$ temperature in ultrahigh vacuum at a base pressure of $2 \times 10^{-10}$ Torr and directly transferred to the STM system (PanScan Freedom, RHK). Chemically etched Pt−Ir tips were used for STM measurement in a constant-current mode at ~10 K. The tips were calibrated on a clean Ag (111) surface. Gwyddion was used for the STM data analysis.